# Amplitude stabilization in a synchronized nonlinear nanomechanical oscillator

Martial Defoort[1,2]*, Sébastien Hentz[2], Steven W. Shaw[3], Oriel Shoshani[4]*

[1]*Univ. Grenoble Alpes, CNRS, Grenoble INP, TIMA, 38000, Grenoble, France*
[2]*Univ. Grenoble Alpes, CEA, LETI, 38000, Grenoble, France*
[3]*Department of Mechanical, and Civil Engineering, Florida Institute of Technology, Melbourne, FL 32901, USA*
[4]*Department of Mechanical Engineering, Ben-Gurion University of the Negev, Beer-Sheva, 84105, Israel*

In contrast to the well-known phenomenon of frequency stabilization in a synchronized noisy nonlinear oscillator, little is known about its amplitude stability. In this paper, we investigate experimentally and theoretically the amplitude evolution and stability of a nonlinear nanomechanical self-sustained oscillator that is synchronized with an external harmonic drive. We show that the phase difference between the tones plays a critical role on the amplitude level, and we demonstrate that in the strongly nonlinear regime, its amplitude fluctuations are reduced considerably. These findings bring to light a new facet of the synchronization phenomenon, extending its range of applications beyond the field of clock-references and suggesting a new means to enhance oscillator amplitude stability.

Synchronization describes the adjustment of rhythms between oscillating objects due to their weak interaction. The synchronization phenomenon has been studied for centuries in various fields of science, tackling issues in both fundamental and applied research. Reported for the first time by Huygens[1] while describing the behavior of two mechanical clocks being progressively in anti-phase regardless of the initials conditions, synchronization has been observed in living systems[2] and social behaviors[3], and is nowadays implemented in modern technology, such as medical applications[4].

In one common occurrence, the synchronization phenomenon is unidirectional, where the frequency of a free-running oscillator is enslaved to that of an external weak perturbation signal. If the external perturbation has a low frequency noise, then this "slave-master" configuration enables one to reduce the frequency noise of the oscillator to that of the perturbation signal[5], such as in pacemakers[4]. This frequency stabilization is an extremely attractive feature for vibrating micro- and nano-electromechanical systems (M/NEMS), which often exhibit relatively strong fluctuations[6–8] due to their small size.

For the past couple of decades, M/NEMS have proven to be key components used in modern technology[9–11], and also useful tools to study physics in both fundamental[12–14] and applied[15–17] domains. Operated as self-sustained oscillators, M/NEMS are also excellent candidates to investigate the synchronization phenomenon, as they have unprecedented control and resolution[18], and are well described by comprehensive models predicting their complex behaviors, such as those arising from Duffing nonlinearities[19]. With these assets, M/NEMS were used to explore synchronization properties, such as mutual synchronization between several oscillators[20], fractional synchronization between oscillators which frequencies share a common divisor[21], and to probe how the Duffing nonlinearity can enhance the frequency locking and phase noise reduction[22–24].

While phase fluctuations of synchronized oscillators have been intensively investigated, the noisy behavior of the amplitude in such systems has been rarely discussed, focusing either theoretically on the linear regime[25] or experimentally on the singular fractional mutual synchronization[26]. In this paper we report both theoretically and experimentally on the effects of synchronization on the amplitude stability of a generic nonlinear nanomechanical oscillator. We begin by presenting our NEMS device and the synchronization parameters involved in this system. Then, we develop theoretical predictions for the phase and amplitude, including the effects of noise, and compare experimentally their evolution within the synchronization range. Finally, we demonstrate that the amplitude fluctuations of the synchronized oscillator can be reduced by increasing the level of its Duffing nonlinearity, down to a level beneath that of the noise of the free-running oscillator.

## Results

**Setup and synchronization range.** The considered NEMS is composed of a silicon-based piezoresistive doubly clamped nanobeam 10 µm long, 160 nm thick and 300 nm width. The resonator's transduction consists of a side electrode from which the applied voltage results in an electrostatic force on the beam, and of a differential piezoresistive readout previously reported[27]. The overall electrical actuation and detection are performed with a lock-in amplifier HF2LI Zurich Instrument. The first in-plane flexural mode of the nanobeam has a natural resonance frequency $f_0 = 27.78$ MHz, a bandwidth $\Delta f = 4.5$ kHz, and a Duffing nonlinear coefficient $\alpha_n = 237$ kHz/mV$^2$ (see Supplementary Information for more details about the calibration, Figs. S1-S2-S3). By means of a phase-locked loop (PLL), the first flexural mode is actuated with a feedback force $F_{osc}$ to operate as a self-sustained free-running oscillator, which is perturbed both by an external tone $F_e$, and an additive noise signal $\xi(t)$. The system is described by the model:

$$\ddot{x} + \Delta\omega\,\dot{x} + \omega_0^2\,x + \frac{8\,\omega_0}{3}\alpha\,x^3 = \frac{F_{osc}}{m}\cos[\Phi_{osc}(t)] + \frac{F_e}{m}\cos[\Phi_e(t)] + \xi(t), \quad (1)$$

with $x$ the displacement of the resonator, $\omega_0 = 2\pi f_0$ its angular eigenfrequency, $\Delta\omega = 2\pi\,\Delta f$ its angular bandwidth (arising from dissipation), $m$ its effective mass, and $\alpha = 2\pi\,\alpha_n$ its angular Duffing coefficient. $\Phi_{osc,e}(t) = \omega_{osc,e}\,t + \varphi_{osc,e}$ describes the phases of the actuation and the external tone. For $F_e = 0$ and $\xi = 0$, Eq. (1) describes the evolution of a driven resonator with an amplitude dependent resonance frequency $\omega_r = \omega_0 + \alpha\,X_0^2$, where $X_0 = F_{osc}/(m\,\omega_0\,\Delta\omega)$ is the operating amplitude[28]. To drive the NEMS as an oscillator at the resonance condition, we set the PLL such that the phase difference between the resonator and the drive is $-\pi/2$ (Fig. 1a), matching the driving frequency with the resonance frequency.

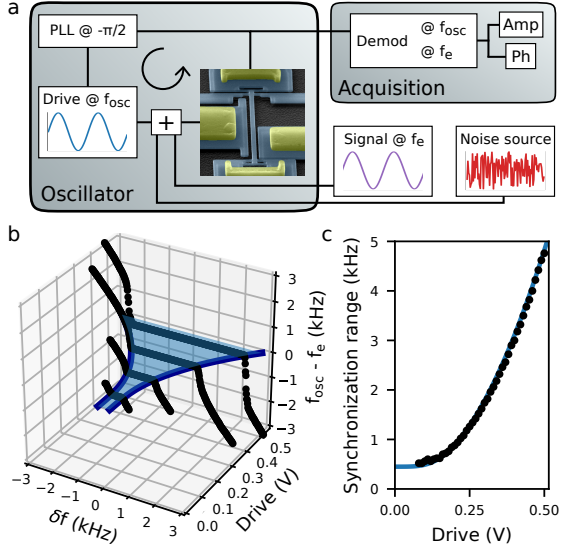

**Figure 1: Experimental setup and synchronization regime. a** The resonator (colored SEM picture of a representative NEMS in inset) is driven as an oscillator at $f_{osc}$ using a feedback loop (arrow), and is subject to an external tone at $f_e$. Using a lock-in amplifier, the output signal of the NEMS is demodulated at either $f_{osc}$ or $f_e$ to obtain amplitude and phase difference. The noise source is turned off for the synchronization range characterization. **b** The oscillator gets synchronized and locked to the frequency $f_e$ for a sufficiently small frequency mismatch $\delta f = f_r - f_e$. **c** The synchronization range increases quadratically with the drive amplitude in the nonlinear regime. The external tone level is set to 10% of the drive. The experimental black data points are plotted on top of the blue theoretical predictions from the model.

The presence of a weak external signal $F_e \ll F_{osc}$, with a phase $\Phi_e$, perturbs the response of the oscillator. By moving to the rotating frame of this external signal in the absence of noise $\xi(t) = 0$ (see supplementary note 1), we obtain the following expressions for the amplitude and phase of the steady-state response:

$$X = X_0 \left(1 - \frac{F_e}{F_{osc}} \sin \delta\varphi\right), \quad (2a)$$

$$\delta\omega = \frac{F_e}{F_{osc}} \left(\frac{\Delta\omega}{2} \cos \delta\varphi + 2\alpha X_0^2 \sin \delta\varphi\right), (2b)$$

where $X$ is the amplitude of the perturbed oscillator, $\delta\varphi = \varphi_X - \varphi_e$ is the phase delay between the oscillator and the perturbation and $\delta\omega = \omega_r - \omega_e$ is the angular frequency difference between the resonance and the perturbation tone.

It follows from Eq. (2b) that the oscillator can be in a steady-state regime even if there is a frequency mismatch between the resonance and the perturbation ($\delta\omega \neq 0$), as long as the right-hand side of Eq. (2b) compensates for it. This is achieved through the phase delay $\delta\varphi$, a free and inner parameter of the system, that balances this mismatch. In practice, as the frequency of the perturbation is detuned from that of the resonance, the phase of the oscillator evolves such that its delay with the perturbation satisfies Eq. (2b). Consequently, the oscillator response remains steady in the rotating frame of the perturbation, which implies that its oscillating frequency is locked on the perturbation tone ($f_{osc} = f_e$), the essential feature of the synchronization phenomenon (Fig. 1b). It also follows from Eq. (2b) that synchronization is possible only for certain values of frequency mismatch, which satisfies the inequality $|\delta\omega| < \Delta\Omega/2$, where the synchronization range $\Delta\Omega$, is given by:

$$\Delta\Omega = \frac{F_e}{F_{osc}} \sqrt{\Delta\omega^2 + (4\alpha X_0^2)^2}. \quad (3)$$

Due to the perturbative nature of the synchronization phenomenon, $F_e$ should remain small compared to $F_{osc}$, and since the dissipation $\Delta\omega$ is usually an intrinsically fixed property of most M/NEMS, the synchronization range can be tuned through the Duffing nonlinearity $\alpha X_0^2$ (Fig. 1c).

**Phase and amplitude behavior in the synchronization regime.** While essential for the synchronization mechanism, the variation of $\delta\varphi$ also affects the amplitude of oscillation [Eq. (2a)], such that the amplitude of a synchronized oscillator also varies with $\delta\omega$ (see supplementary note 1). In the linear regime ($\alpha X_0^2 \ll \Delta\omega$), a perfect frequency matching between the free-running oscillator and the external tone ($\delta\omega = 0$) induces a phase delay of $-\pi/2$, readily deduced from Eq. (2b) and observed experimentally (Fig. 2). This phase delay of $-\pi/2$ is identical to the phase delay of the PLL at the resonance amplitude, such that the driving and the perturbation signals linearly add (Fig. 3). However, a deviation from the center of the synchronization range induces a parabolic variation in the amplitude of the oscillator, directly arising from the sinusoid in Eq. (2a) for $\delta\varphi \approx -\pi/2$. As the oscillator enters the nonlinear regime, the evolution of both the phase and amplitude as functions of the frequency detuning become more complex and asymmetric. This behavior is a direct consequence of the amplitude-to-frequency conversion arising from the backbone curve of the Duffing oscillator, which leads to the extra $\sin \delta\varphi$ term in Eq. (2b).

Deep in the nonlinear regime ($\alpha X_0^2 \gg \Delta\omega$), this Duffing term becomes predominant, such that a perfect frequency matching between the free-running oscillator and the external tone ($\delta\omega = 0$) corresponds to a zero-phase delay $\delta\varphi = 0$, as can be seen both from the experimental measurements (Fig. 2b), and from Eq. (2b). Consequently, the amplitude of the oscillator is not affected by the strength of the perturbation (Eq. (2a)) and remains equal to the amplitude of the free-running oscillator regardless of the amplitude of the external tone $F_e$ (Fig. 3b). As the frequency of the external tone is detuned from that of the free-running oscillator, the former parabolic behavior of the amplitude evolves to a linear dependency (Fig. 3c). Since the Duffing nonlinearity $\alpha$ is positive for the present device, the amplitude of the synchronized oscillator becomes the largest (smallest, resp.) at the negative (positive, resp.) boundaries of the synchronization range (Fig. 3a).

The frequency and phase fluctuations in such systems have been intensively studied both theoretically and experimentally[20,24]. In light of Eq. (2a), it should be noted that since the amplitude of a synchronized oscillator depends on $\delta\varphi$ and hence $\delta\omega$, these phase fluctuations have a direct impact on the oscillator's amplitude stability, which may reduce the range of applications of the synchronization phenomenon[29]. However, as the nonlinearity of the oscillator increases, this frequency-to-amplitude conversion decreases (Fig. 3c) thereby reducing the impact of frequency fluctuations on the amplitude stability of the synchronized oscillator.

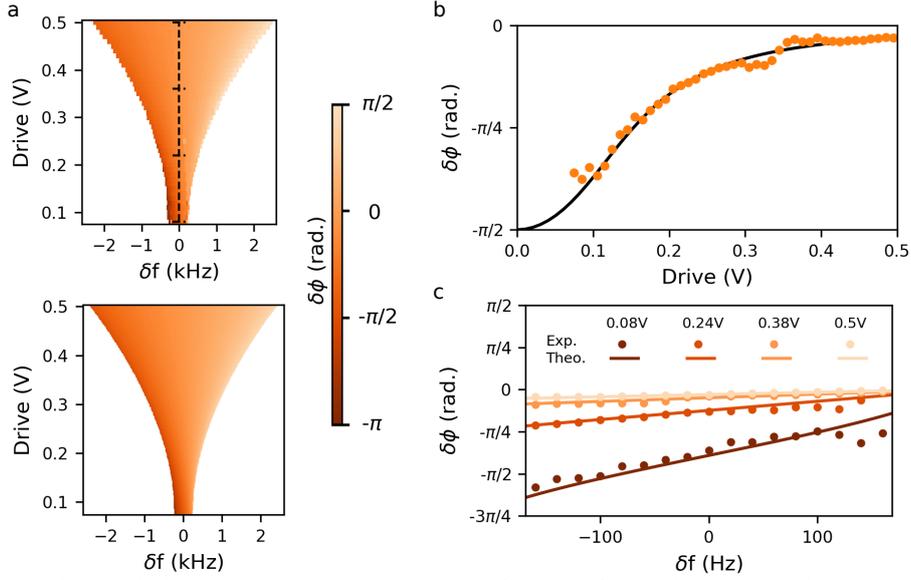

**Figure 2: Phase variation within the synchronization range as a function of the drive $F_{osc}$, keeping the ratio between the perturbation and the drive $F_e/F_{osc}$ at a constant level of 10%. a** As the tone of the perturbation is detuned from the frequency of the free running oscillator, their phase difference $\delta\varphi$ adjusts to maintain synchronization (top panel: experimental results, bottom panel: model). **b** Cross-section of panel (a) along $\delta f = 0$, where the phase delay between the oscillator and the perturbation shrinks as the system enters the nonlinear regime (line: theory, dots: experiment). **c** Cross-sections of panel (a) along different drive levels near zero detuning, where the phase delay varies less with the increasing Duffing nonlinearity.

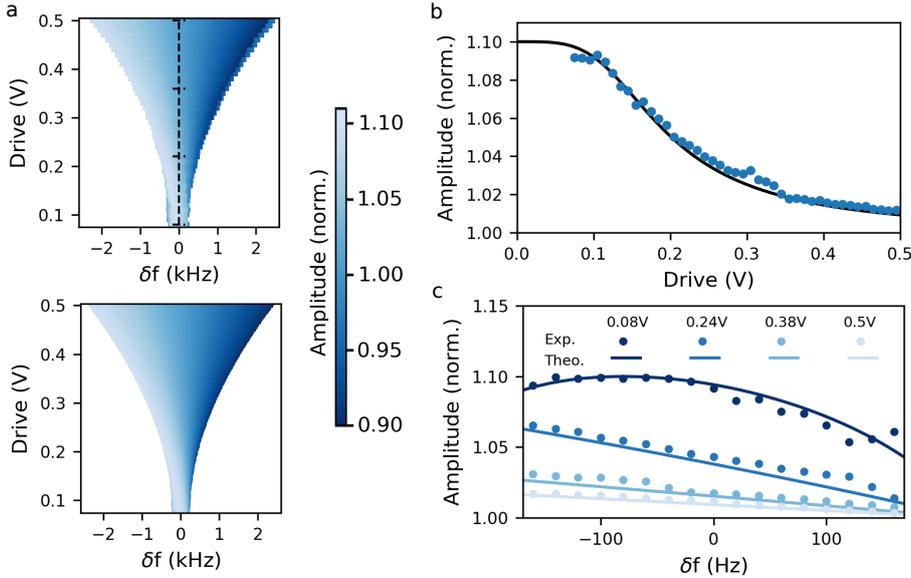

**Figure 3: Amplitude variation within the synchronization range as a function of the drive $F_{osc}$, keeping the ratio between the perturbation and the drive $F_e/F_{osc}$ at a constant level of 10%.** The amplitude is normalized to that of the free-running oscillator for the same drive ($X/X_0$). **a** The phase delay induced by the frequency detuning leads to an amplitude variation (top panel: experimental results, bottom panel: model). **b** Cross-section of panel (a) along $\delta f = 0$, where the amplitude drops towards the free-running oscillator amplitude as the nonlinearity increases (line: theory, dots: experiment). **c** Cross-sections of panel (a) along different drive levels near zero detuning, where the amplitude variation changes from parabolic to linear with a decreasing slope.

**Amplitude stabilization in the nonlinear regime.** As opposed to the frequency fluctuations, the influence of the amplitude fluctuations of the free-running oscillator on those of the synchronized oscillator has been so far overlooked. To quantitatively investigate these fluctuations in the synchronization regime, we inject to the oscillator an additive noise signal generated by a Siglent SDG1032X voltage source (Fig. 1), introduced as $\xi(t)$ in Eq. (1). This noise is assumed small, with a zero mean, and a correlation time $\tau_\xi = \langle\xi^2\rangle^{-1}\int_0^\infty \langle\xi(t)\xi(t+\tau)\rangle d\tau$ that is significantly smaller than the relaxation time of the oscillator $\tau_r = 1/\Delta\omega$. Thus, we apply the method of stochastic averaging and linearize the resulting stochastic equations of the amplitude and the phase delay with respect to deterministic operating point $(X, \delta\varphi)$. This procedure (supplementary note 1) leads to a pair of linear coupled Langevin equations from which we calculate the power spectral density $S_{u_X}(\omega_s) = \delta X(\omega_s)^2$, with $\delta X$ the amplitude fluctuations of the synchronized oscillator and $\omega_s$ is the offset frequency from the carrier frequency $\omega_e$. Focusing on a perfect frequency match ($\delta\omega = 0$), these amplitude fluctuations fall back to the standard Lorentzian spectral density of the free-running oscillator in the linear regime $\delta X_0$. However, deep in the nonlinear regime ($\alpha X_0^2 \gg \Delta\omega$), the power spectral density of the amplitude fluctuations reduces to:

$$S_{u_X}^{nonlin}(\omega_s) = \frac{\left[\left(\frac{F_e}{F_{osc}}\right)^2 \Delta\omega^2 + 4\omega_s^2\right] S_\xi(\omega_e)}{\omega_e^2 \left[\omega_s^2 \Delta\omega^2 + 4\left(\frac{F_e}{F_{osc}} \Delta\omega \alpha X_0^2 - \omega_s^2\right)^2\right]}, \quad (4)$$

where $S_\xi(\omega_e)$ is the noise intensity of the source at the carrier frequency. Two main features arise from this nonlinear regime. First, the spectral density is peaked at an offset frequency $\omega_{s_{peak}} = \sqrt{F_e \Delta\omega \alpha X_0^2 / F_{osc}}$ from the carrier frequency $\omega_e$. Second, near the carrier frequency ($\omega_s = 0$), the amplitude fluctuations reduce as the nonlinearity increases, dropping below that of the free-running oscillator, following $\frac{\delta X}{\delta X_0} = \frac{\Delta\omega}{4\alpha X_0^2}$ (supplementary note 1). To explore these behaviors experimentally, we probed the oscillator's response at the resonance frequency in both free-running and synchronized regimes with a fixed input noise amplitude as the system enters the nonlinear regime. We find a quantitative agreement with the theoretical predictions (Fig. 4), resulting in a decrease of the amplitude noise by a factor four near the carrier frequency (Fig. 4b). This substantial noise-reduction was performed with a relatively small Duffing nonlinearity, less than three bandwidths (Fig. S1), easily accessible to most micro/nanomechanical resonators.

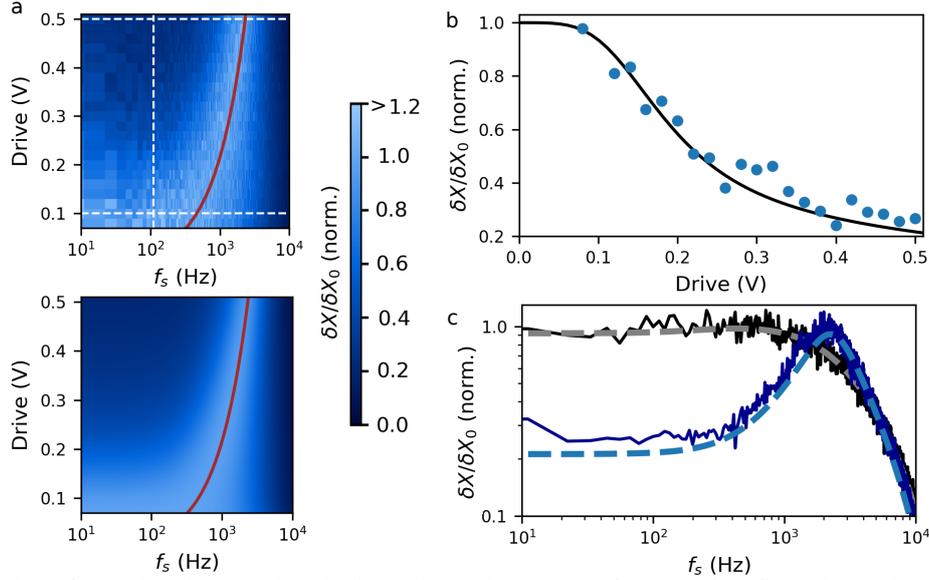

**Figure 4: Amplitude fluctuations in the synchronization regime at the resonance frequency for a fixed added noise of 0.5 Vstd.** The amplitude fluctuations are normalized to that of the free-running oscillator for the same drive levels (see Fig. S4 and supplementary note 2). **a** Spectral density of the amplitude fluctuations as the nonlinearity of the oscillator increases (top panel: experimental results, bottom panel: model). The red line shows the theoretical position of the peak at $f_{s_{peak}}$. **b** Cross-section of panel (a) near the carrier frequency, highlighting the reduction in amplitude fluctuations as the oscillator enters the nonlinear regime (black line: theory, blue dots: experimental results). The spectral frequency was purposely shifted by ~100 Hz from the carrier frequency to avoid 1/f noise. **c** Cross-section of panel (a) along 0.1 V (black, linear regime) and 0.5 V (blue, Duffing regime), presenting the evolution of the spectral density from a regime similar to that of a free-running oscillator to the nonlinear regime where the fluctuations are shifted away from the carrier frequency (dashed line: theory, continuous line: experiment).

Qualitatively, we can explain the source of the amplitude noise-reduction from a deterministic analysis by considering the additive noise as an amplitude perturbation in the rotating frame approximation. Such an amplitude variation generates a frequency shift due to the amplitude-to-frequency conversion of the Duffing nonlinearity. However, this frequency shift is compensated by a phase delay adjustment due to the synchronization regime (Fig. 2), which acts back on the amplitude of the oscillator (Fig. 3). In an initially perfect frequency matching ($\delta\omega = 0$) deep in the nonlinear regime ($\alpha X_0^2 \gg \Delta\omega$), this retroaction tends toward an exact compensation of the initially added amplitude perturbation. Going a step further in this deterministic approach, the dynamics around the stable synchronized solution reveals that the Duffing nonlinearity acts as an effective restoring force that bounds the motion of the amplitude fluctuations, thereby reducing their impact on the oscillator's amplitude (supplementary note 1).

## Discussion

The frequency locking property of the synchronization phenomenon is ideal when it comes to reducing the frequency fluctuations of an oscillator, but it inherently requires the master signal to be cleaner than the synchronized oscillator. On the other hand, the reduction of amplitude fluctuations is not a locking mechanism, it is directly related to the oscillator's nonlinear properties and does not involve strong requirements on the amplitude fluctuations of the master signal. In both cases, this noise reduction prevents the use of synchronization for sensing applications, as the sensing mechanism is thereby reduced. However, the amplitude stabilization property could have a substantial impact for resonant micro/nano-actuators. For ultrasound transducers, the interaction with the environment (gas or liquids) drastically damps the acoustic pressure level[30]. Achieving large amplitudes is therefore essential, which is usually performed with arrays of transducers, and improving their resolution through synchronization could open new perspectives for airborne communication schemes. In the case of mechanical vibratory rate gyroscopes[31], the amplitude of the actuation mode is traditionally stabilized with a proportional integrator (PI) loop controller to enhance the angular rate sensitivity with an improved signal-to-noise ratio. However, as the amplitude of the actuation mode increases, the resonator enters the Duffing regime, such that a direct control on the amplitude might induce frequency fluctuations, thereby reducing the sensor's performance, which would be avoided with this nonlinear synchronization regime. Finally, micro/nano-mechanical resonators have also demonstrated logic gate and memory applications for digital implementation[16,32]. In this context, amplitude stabilization would enhance the resolution for amplitude-based digital encoding such as quadrature amplitude modulation (QAM).

In conclusion, we demonstrated both experimentally and theoretically that the phase delay between the oscillator and the external tone plays a crucial role in the amplitude level of the synchronized oscillator. Near frequency matching, the impact of the external signal on the amplitude fades as the Duffing nonlinearity of the oscillator increases. This behavior is followed by a reduction of the amplitude fluctuations of the system to a level below that of the free-running oscillator. Our study explores the largely ignored amplitude stabilization property of the synchronization phenomenon so far ignored and paves the way to implement synchronization in drastically different applications, exploiting the amplitude stabilization rather than the frequency locking mechanism.

## Methods

**Synchronization regime.** The synchronization range of the oscillator extends far from the resonance frequency of the free-running oscillator. However, it is necessary to first match the external tone to the frequency of the oscillator to enter in the synchronization regime. Starting from that working point, it is then possible to explore the synchronization phenomenon as a function of the frequency detuning.

**Amplitude and phase measurements.** Oscillators suffer from frequency noise, which directly impacts the estimated frequency detuning from the resonance frequency. When characterizing the amplitude and phase within the synchronization regime, it is essential to average over several independent measurements, each of them comprising:

1 - measuring the resonance frequency
2 - turning on the external tone
3 - entering the synchronization regime
4 - applying the desired frequency detuning
5 - measuring the synchronization state
6 - turning off the external tone

Depending on the frequency fluctuations of each oscillator, steps 4 and 5 may be looped before reinitiating the procedure. The experimental results in Fig. 2 and Fig. 3 are the result of an average over 17 measurements.

**Amplitude fluctuations measurements.** The experimental results presented in Fig.4 are the result of an average over 40 spectra for each driving amplitude, to obtain good resolution of the amplitude fluctuations.

## Data availability

The data that support the findings of this study are available from the corresponding authors upon reasonable request.

## Author contributions

M.D. performed the experiments and the data analysis. O.S. performed the theoretical analysis. S.H. and S.S. supervised the experimental and theoretical parts, respectively. All authors co-wrote the manuscript.

## Competing interests

The authors declare no competing interests.

## Additional information

**Correspondence** should be addressed to M.D. or O.S at martial.defoort@univ-grenoble-alpes.fr or oriels@bgu.ac.il